\documentclass[journal,draftcls,onecolumn,12pt,twoside]{IEEEtranTCOM}
\normalsize
\usepackage{setspace}
\usepackage{amsmath,amssymb,amsthm}
\usepackage{graphicx}
\usepackage{algorithmic,algorithm}
\usepackage{subfigure}
\usepackage{rotating}
\usepackage{cite}

\newcommand{\twotriangle}{\hfill $\bigtriangleup \bigtriangleup$  }
\newcommand{\eax}{\twotriangle  \end{example}}

\floatname{algorithm}{Algorithm}

\newtheorem{Theorem}{Theorem}
    
    \newtheorem{lemma}[Theorem]{Lemma}
    
    \theoremstyle{definition}
        
        \newtheorem{example}{Example}

        \newtheorem{Algorithm}{Algorithm}

    \theoremstyle{remark}

    \newenvironment{Proof}{\par\noindent{\sc Proof}\quad}{\hfill\qed\par\smallskip}

\begin{document}

\title{A Simplified Min-Sum Decoding Algorithm for Non-Binary LDPC Codes}
\author{Chung-Li (Jason) Wang, Xiaoheng Chen, Zongwang Li, and Shaohua Yang
\thanks{Chung-Li (Jason) Wang, Zongwang Li, and Shaohua Yang are with LSI Corporation,
Milpitas, CA 95035, USA
(e-mail: \{ChungLi.Wang, Zongwang.Li, Shaohua.Yang\}@lsi.com); Xiaoheng Chen was with Microsoft Corporation, Redmond, WA 98052, USA. He is now with Sandisk Corporation, Milpitas, CA 95035, USA (e-mail: chen.xiaoheng@gmail.com).}
}
\markboth{IEEE Transactions on Communications}%
{Submitted paper}
\maketitle

\begin{abstract}
Non-binary low-density parity-check codes are robust to
various channel impairments. However, based on the existing decoding algorithms, the decoder implementations are expensive because of their excessive computational complexity and memory
usage. Based on the combinatorial optimization, we present an approximation method for the check node processing. The simulation results demonstrate that our scheme has small performance loss over the additive white Gaussian noise channel and independent Rayleigh fading channel. Furthermore, the proposed reduced-complexity realization provides significant savings on hardware, so it yields a good performance-complexity tradeoff and can be efficiently implemented.
\end{abstract}

\IEEEpeerreviewmaketitle

\section{Introduction}
\label{sec:intro} Binary low-density parity-check (LDPC) codes,
discovered by Gallager in 1962~\cite{gallager1962ldp}, were
rediscovered and shown to approach Shannon capacity in the late
1990s \cite{mackay1999gec}.  Since their rediscovery, a great deal
of research has been conducted in the study of code construction methods, decoding techniques, and performance analysis.  With
hardware-efficient decoding algorithms such as the min-sum
algorithm \cite{fossorier1999rci}, practical decoders can be implemented for effective error-control. Therefore, binary LDPC codes have been considered for a wide
range of applications such as satellite broadcasting, wireless
communications, optical communications, and high-density storage
systems.

As the extension of the binary LDPC codes over the Galois field of order $q$, non-binary LDPC (NB-LDPC)
codes, also known as $q$-ary LDPC codes, were first
investigated by Davey and MacKay in 1998~\cite{davey2002low}. They
extended the sum-product algorithm (SPA) for binary LDPC codes
to decode $q$-ary LDPC codes and referred to this extension as the
$q$-ary SPA (QSPA). Based on the fast Fourier transform (FFT), they
devised an equivalent realization called FFT-QSPA to reduce the
computational complexity of QSPA for codes with $q$ as a power of
2~\cite{davey2002low}. With good construction methods
\cite{bennatan2006design,song2006algebraic,zeng2008construction,zhou2009high,zhou2009construction},
NB-LDPC codes decoded with the FFT-QSPA outperform Reed-Solomon
codes decoded with the algebraic soft-decision Koetter-Vardy
algorithm \cite{koetter2003algebraic}.

As a class of capacity approaching codes, NB-LDPC codes are capable
of correcting symbol-wise errors and have recently
been actively studied by numerous researchers.  However, despite the
excellent error performance of NB-LDPC codes, very little research
contribution has been made for VLSI decoder implementations due to the lack of
hardware-efficient decoding algorithms.  Even though the FFT-QSPA
significantly reduces the number of computations for the QSPA, its
complexity is still too high for practical applications, since it
incorporates a great number of multiplications in probability domain
for both check node (CN) and variable node (VN) processing.
Thus logarithmic domain approaches were developed to
approximate the QSPA, such as the extended min-sum algorithm (EMSA),
which applies message truncation and sorting to further reduce
complexity and memory requirements
\cite{declercq2007decoding,voicila2007architecture}.  The second
widely used algorithm is the min-max algorithm (MMA) \cite{min:max},
which replaces the sum operations in the CN processing by max
operations. With an optimal scaling or offset factor, the EMSA and
MMA can cause less than 0.2 dB performance loss in terms of
signal-to-noise ratio (SNR) compared to the QSPA.  However,
implementing the EMSA and MMA still requires excessive silicon area,
making the decoder considerably expensive for practical designs
\cite{wang:tcas1,zhang:tcas1,zhang:tvlsi,Xiaoheng:TCASI:Nonbinary}.
Besides the QSPA and its approximations, two reliability-based
algorithms were proposed towards much lower complexity based on the
concept of simple orthogonal check-sums used in the one-step
majority-logic decoding \cite{chen2010two}. Nevertheless, both
algorithms incur at least 0.8 dB of SNR loss compared to the
FFT-QSPA.  Moreover, they are effective for decoding only when the
parity-check matrix has a relatively large column weight.
Consequently, the existing decoding algorithms are either too costly
to implement or only applicable to limited code classes at cost of
huge performance degradation.

Therefore, we propose a reduced-complexity decoding algorithm,
called the simplified min-sum algorithm (SMSA), which is derived
from our analysis of the EMSA based on the combinatorial
optimization. Compared to the QSPA, the SMSA shows small SNR loss, which is similar to that of the EMSA and MMA. Regarding
the complexity of the CN processing, the SMSA saves
around 60\% to 70\% of computations compared to the EMSA.  Also, the SMSA
provides an exceptional saving of memory usage in the decoder
design. According to our simulation results and complexity
estimation, this decoding algorithm achieves a favorable tradeoff
between error performance and implementation cost.

The rest of the paper is organized as follows.  The NB-LDPC code and
EMSA decoding are reviewed in Section \ref{sec:code}.  The SMSA is
derived and developed in Section \ref{sec:cn-processing}.  The error
performance simulation results are summarized in Section
\ref{sec:simulation}.  In Section \ref{sec:complexity}, the SMSA is
compared with the EMSA in terms of complexity and memory usage.  At last, Section \ref{sec:conclusion} concludes
this paper.

\section{NB-LDPC Codes and Iterative Decoding}
\label{sec:code}

Let $\text{GF}(q)$ denote a finite field of $q$ elements with
addition $\oplus$ and multiplication $\otimes$. We will focus on the
field with characteristic $2$, i.e., $q=2^p$. In such a field, each
element has a binary representation, which is a vector of $p$ bits
and can be translated to a decimal number. Thus we label the
elements in $\text{GF}(2^p)$ as $\{0,1,2,\ldots 2^p-1\}$. An $(n,r)$
$q$-ary LDPC code \emph{C} is given by the null space of an $m\times
n$ sparse parity-check matrix $\mathbf{H}=[h_{i,j}]$ over
$\text{GF}(q)$, with the dimension $r$.

The parity-check matrix $\mathbf{H}$ can be represented graphically
by a Tanner graph, which is a bipartite graph with two disjoint
variable node (VN) and check node (CN) classes. The $j$-th VN
represents the $j$-th column of $\mathbf{H}$, which is associated
with the $j$-th symbol of the $q$-ary codeword. The $i$-th CN
represents its $i$-th row, i.e., the $i$-th $q$-ary parity check of
$\mathbf{H}$. The $j$-th VN and $i$-th CN are connected by an edge
if $h_{i,j}\neq 0$. This implies that the $j$-th code symbol is
checked by the $i$-th parity check. Thus for $0 \leq i < m$ and $0
\leq j < n$, we define $N_i = \{j: 0\leq j<n, h_{i,j} \neq 0\}$, and
$M_j = \{i: 0\leq i< n, h_{i,j} \neq 0\}$. The size of $N_i$ is
referred to as the CN degree of the $i$-th CN, denoted as $|N_i|$.
The size of $M_j$ is referred to as the VN degree of the $j$-th VN,
denoted as $|M_j|$. If both VN and CN degrees are invariable,
letting $d_v=|M_j|$ and $d_c=|N_i|$, such a code is called a
$(d_v,d_c)$-regular code. Otherwise it is an irregular code.

Similarly as binary LDPC codes, $q$-ary LDPC codes can be decoded
iteratively by the message passing algorithm, in which messages are
passed through the edges between the CNs and VNs. In the QSPA, EMSA,
and MMA, a message is a vector composed of $q$ sub-messages, or
simply say, entries. Let
${\lambda}_j=[{\lambda}_j(0),{\lambda}_j(1),\ldots,{\lambda}_j(q-1)]$
be the \emph{a priori} information of the $j$-th code symbol from
the channel.  Assuming that $X_j$ is the $j$-th code symbol, the
$d$-th sub-message of ${\lambda}_j$ is a log-likelihood reliability (LLR) defined
as ${\lambda}_j(d)=\log(\text{Prob}(X_j=z_j)/\text{Prob}(X_j=d))$.
$z_j$ is the most likely (ML) symbol for $X_j$, i.e.,
$z_j=\arg\max_{d\in\text{GF}(q)} \text{Prob}(X_j=d)$, and ${\bf z}=[z_j]_{j=1\ldots n}$.
The smaller ${\lambda}_j(d)$ is, the more likely $X_j=d$ is.
Let ${\alpha}_{i,j}$ and ${\beta}_{i,j}$ be the VN-to-CN (V2C) and
CN-to-VN (C2V) soft messages between the $i$-th CN and $j$-th VN
respectively. For all $d\in \text{GF}(q)$, the $d$-th entry of
${\alpha}_{i,j}$, denoted as ${\alpha}_{i,j}(d)$, is the logarithmic
reliability of $d$ from the VN perspective. ${a}_{i,j}$ is the
symbol with the smallest reliability, i.e., the ML symbol of the V2C message. With ${x}_{i,j}=X_j\otimes
h_{i,j}$, we let
${\alpha}_{i,j}(d)=\log(\text{Prob}(x_{i,j}=a_{i,j})/\text{Prob}(x_{i,j}=d))$
and ${\alpha}_{i,j}(a_{i,j})=0$. ${b}_{i,j}$ and ${\beta}_{i,j}(d)$
are defined from the CN perspective similarly. The EMSA can be
summarized as follows.
\begin{Algorithm}\label{algorithm:emsa}
\noindent\textbf{The Extended Min-Sum Algorithm}\\
\emph{Initialization}: Set $z_j = \arg\min_{d\in
\text{GF}(q)}{\lambda}_j(d)$. For all $i,j$ with $h_{i,j}\neq 0$, set $\alpha_{i,j}(h_{i,j}\otimes d) =
{\lambda}_j(d)$. Set $\kappa=0$.
\begin{itemize}
\item Step 1) Parity check: Compute the syndrome ${\bf z}\otimes {\bf
H}^{\text{T}}$. If ${\bf z}\otimes {\bf H}^{\text{T}} = {\bf
0}$, stop decoding and output ${\bf z}$ as the decoded codeword;
otherwise go to Step 2.
\item Step 2) If $\kappa = \kappa_{\text{max}}$, stop decoding and declare a decoding failure; otherwise, go to Step 3.

\item Step 3) \emph{CN processing:} Let the configurations $\mathcal{L}_i({x}_{i,j}=d)$ be the sequence $[x_{i,j'}]_{j'\in N_i}$ such that $\sum^{\oplus}_{j'\in N_i}x_{i,j'}=0$ and $x_{i,j}=d$. With a preset scaling factor $0< c\leq 1$, compute the C2V messages by
 \begin{equation}\label{eq:orig-min-sum}{\beta}_{i,j}(d)=c\cdot\min_{\mathcal{L}_i({x}_{i,j}=d)}\sum_{j'\in N_i\setminus j}{\alpha}_{i,j'}({x}_{i,j'}).\end{equation}
\item Step 4) \emph{VN processing:} $\kappa\leftarrow\kappa+1$. Compute V2C messages in two steps. First compute the primitive messages by
    \begin{equation}\label{eq:var-proc}\hat{\alpha}_{i,j}(h_{i,j}\otimes d)=\lambda_j(d)+\sum_{i'\in M_j\setminus i}{\beta}_{i',j}(h_{i',j}\otimes d). \end{equation}
    \item Step 5) \emph{Message normalization}: Obtain V2C messages by normalizing with respect to the ML symbol
\begin{equation}\label{eq:var-hd}{a}_{i,j}=\arg\min_{d\in\text{GF}(q)}\hat{\alpha}_{i,j}(d).\end{equation}
    \begin{equation}\label{eq:var-norm}{\alpha}_{i,j}(d)=\hat{\alpha}_{i,j}(d)-\hat{\alpha}_{i,j}({a}_{i,j}).\end{equation}
\item Step 6) \emph{Tentative Decisions:}
 \begin{equation}\label{eq:hd-1}\hat{\lambda}_j(d)={\lambda}_j(d)+\sum_{i\in M_j}{\beta}_{i,j}(h_{i,j}\otimes d). \end{equation}
  \begin{equation}\label{eq:hd-2}{z}_j=\arg\min_{d\in\text{GF}(q)}\hat{\lambda}_j(d).\end{equation}
  \item Go to Step 1.
\end{itemize}
\end{Algorithm}
\section{A Simplified Min-Sum Decoding Algorithm}
\label{sec:cn-processing}
In this section we develop the simplified min-sum decoding algorithm. In the first part, we analyze the configurations and propose the approximation of the CN processing. Then in the second part,  a practical scheme is presented to achieve the tradeoff between complexity and performance.

\subsection{Algorithm Derivation and Description}\label{sec:SMSA}

In the beginning, two differences between the SMSA and EMSA are
introduced. First, the SMSA utilizes ${a}_{i,j}$ (${b}_{i,j}$) as
the V2C (C2V) hard message, which indicates the ML symbol given by the V2C (C2V) message. Second, the
reordering of soft message entries in the SMSA is defined as:
\begin{flalign}\tilde{\alpha}_{i,j}(\delta)&={\alpha}_{i,j}(\delta\oplus a_{i,j})\\
\tilde{\beta}_{i,j}(\delta)&={\beta}_{i,j}(\delta\oplus
b_{i,j}),\end{flalign} for all $i,j$ with $h_{i,j}\neq 0$. While  in the
EMSA the
arrangement of entries is made by the \emph{absolute} value, the SMSA arranges the entries by the \emph{relative} value to
the hard message, expressed and denoted as the deviation $\delta$.
Thus before the CN processing of the SMSA, the messages are required
to be transformed from the absolute space to the deviation space.

Equation (\ref{eq:orig-min-sum}) performs the combinatorial
optimization over all configurations. If we regard the sum of reliabilities $\sum_{j'\in N_i\setminus j}{\alpha}_{i,j'}({x}_{i,j'})$ as the reliability of the configuration $[x_{i,j'}]_{j'\in N_i}$, this operation actually provides the \emph{most likely} configuration and assigns its reliability to the result. However, the size of its search space
is of $O(q ^{d_c})$ and leads to excessive complexity. Fortunately, in \cite{declercq2007decoding} it is observed that
the optimization tends to choose the configuration with more entries equal to the V2C hard messages.
Therefore, if we define the order
as the number of all $j'\in N_i\setminus j$ such that
${x}_{i,j'}\neq {a}_{i,j'}$, (\ref{eq:orig-min-sum}) can be reduced by
utilizing the order-$k$ subset, denoted as
$\mathcal{L}_i^{(k)}({x}_{i,j}=d)$, which consists of the
configurations of orders not higher than $k$. Limiting the size of the search space gives a reduced-search algorithm with performance loss \cite{declercq2007decoding}, so adjusting $k$ can be used to give a tradeoff between performance and complexity. We denote the order-$k$ C2V soft
message by ${\beta}^{(k)}$ (with the subscript $i,j$ omitted for
clearness), i.e.
\begin{equation}\label{eq:approx1-min-sum}
{\beta}_{i,j}(d)\leq{\beta}^{(k)}(d)=\min_{ \mathcal{L}_i^{(k)}({x}_{i,j}=d)}\sum_{j'\in N_i\setminus j}{\alpha}_{i,j'}({x}_{i,j'}),
\end{equation}
since
$\mathcal{L}_i^{(k)}({x}_{i,j}=d)\subseteq\mathcal{L}_i({x}_{i,j}=d)$.
In the following context, we will show the computations for the hard
message and order-$1$ soft message. Then these messages will be used
to generate high-order messages. The hard message is simply given by
Theorem 1.

\begin{Theorem}\label{thm:1}
The hard message ${b}_{i,j}$ is determined by
\begin{equation}\label{chk-proc-hd}
{b}_{i,j}\equiv\arg\min_{d\in\text{GF}(q)} \beta_{i,j}(d)={\sum_{j'\in N_i\setminus j}}^\oplus{a}_{i,j}.
\end{equation}
Besides, for any order $k$, ${\beta}_{i,j}({b}_{i,j})=\tilde{\beta}_{i,j}(0)={\beta}^{(k)}({b}_{i,j})=0$.
\end{Theorem}
\begin{Proof}
From (\ref{eq:approx1-min-sum}) the inequality is obtained as:
\begin{equation}\label{eq:sum_of_min}{\beta}^{(k)}(d)\geq \sum_{j'\in N_i\setminus j}\min_{{x}_{i,j'}\in\text{GF}(q)}{\alpha}_{i,j'}({x}_{i,j'})= \sum_{j'\in N_i\setminus j}{\alpha}_{i,j'}(a_{i,j'}).   \end{equation}
If ${x}_{i,j}=b_{i,j}$ and ${x}_{i,j'}=a_{i,j'}$ for all $j'\in N_i\setminus j$, we get an order-$0$  configuration, included in $\mathcal{L}_i^{(k)}({x}_{i,j}=b_{i,j})$ for any $k$. Thus one can find that the equation (\ref{eq:sum_of_min}) holds if $d=b_{i,j}$, and ${\beta}^{(k)}(b_{i,j})$ has the smallest reliability. It follows that for any $k$
\begin{equation}\label{eq:sum_of_min2}{\beta}_{i,j}(b_{i,j})={\beta}^{(k)}(b_{i,j})= \sum_{j'\in N_i\setminus j}{\alpha}_{i,j'}(a_{i,j'})=0.    \end{equation}
\end{Proof}

Based on Theorem \ref{thm:1}, for any $k$ we can define  the
order-$k$ message
$\tilde{\beta}^{(k)}(\delta)={\beta}^{(k)}(\delta\oplus{b}_{i,j})$
in the deviation space. For $\delta\neq 0$, the order-$1$ C2V
message $\tilde{\beta}^{(1)}(\delta)$  can be determined by Theorem
\ref{lm:ord-1}, which performs a combinatorial optimization in the
deviation space.

\begin{Theorem}\label{lm:ord-1}
With $\delta= {b}_{i,j}\oplus d$, the order-$1$ soft message is determined by
\begin{equation}\begin{split}\label{eq:chk-message-ord1}&{\beta}^{(1)}(d)=\tilde{\beta}^{(1)}(\delta)\\&=\min_{\small j''\in N_i\setminus j}\bigg( \sum_{\small j'\in N_i\setminus \{j,j''\}} {\alpha}_{i,{j'}}({a}_{i,{j'}})+{\alpha}_{i,{j''}}({a}_{i,{j''}}\oplus \delta)\bigg)\\&=\min_{\small j''\in N_i\setminus j}\tilde{\alpha}_{i,{j''}}(\delta).  \end{split}\end{equation}
\end{Theorem}
\begin{Proof}
  According to the definition of the order, each configuration in $\mathcal{L}_i^{(1)}({x}_{i,j}=d)$ has  ${x}_{i,j''}={a}_{i,{j''}}\oplus \delta$ for some $j''\in N_i\setminus j$ and ${x}_{i,j'}={a}_{i,{j'}}$ for all $j'\in N_i\setminus \{j,j''\}$, since $d\oplus({a}_{i,{j''}}\oplus \delta)\oplus\sum^{\oplus}_{j'\neq \{j,j''\}}{a}_{i,{j'}}=0$. It follows that selecting $j''\in N_i\setminus j$ is equivalent to selecting an order-$1$ configuration in $\mathcal{L}_i^{(1)}({x}_{i,j}=d)$. Correspondingly, minimizing $\tilde{\alpha}_{i,{j''}}(\delta)$ over $j''$ in the deviation space is equivalent to minimizing ${\alpha}_{i,{j''}}({a}_{i,{j''}}\oplus \delta)$ over the configurations in the absolute space. Hence searching for $j''$ to minimize the sum in the bracket of (\ref{eq:chk-message-ord1}) yields  ${\beta}^{(1)}(d)$.
\end{Proof}

Similarly to Theorem $\ref{lm:ord-1}$, in the absolute space
an order-$k$  configuration  can be determined by assigning a deviation to each of $k$ VNs selected from $N_i\setminus j$, i.e., $x_{i,j'}=\delta_{j'}\oplus a_{i,j'}$ with $\delta_{j'}\neq 0$ for selected VNs and $\delta_{j'}=0$ for all other VNs. Thus in the deviation space, the order-$k$ message can be computed as follows:
\begin{Theorem}\label{lm:ord-k}
With $\delta= {b}_{i,j}\oplus d$, choosing a combination of $k$ symbols from $\text{GF}(q)$ (denoted as $\delta_1\ldots \delta_k$) and picking a permutation of $k$ different VNs  from the set $N_i\setminus j$ (denoted as $j_1, j_2,\ldots, j_k$), the order-$k$ soft message is given by
\begin{equation}\label{eq:norm-high-order}
{\beta}^{(k)}(d)=\tilde{\beta}^{(k)}(\delta)=\min_{{\sum^\oplus}_{\ell=1}^{k} \delta_\ell= \delta }\min_{\begin{subarray}{c}  j_1,\ldots,j_k\in N_i\setminus j \\ j_1\neq \ldots\neq j_k \end{subarray}}\sum_{\ell=1}^{k}\tilde{\alpha}_{i,{j_\ell}}(\delta_\ell).
\end{equation}
\end{Theorem}
Theorem \ref{lm:ord-k} shows that  the configuration set can be analyzed as the Cartesian product of the set of symbol combinations and that of VN permutations. For Equation (\ref{eq:norm-high-order}) the required set of combinations can be generated
according to Theorem \ref{thm:exclude-rep}.
\begin{Theorem}\label{thm:exclude-rep}
The set of $k$-symbol combinations $\delta_1\ldots \delta_k$ for (\ref{eq:norm-high-order}) can be obtained by choosing $k$ symbols from $\text{GF}(q)$ of which there exists no subset with the sum equal to $0$.
\end{Theorem}
\begin{Proof}
Suppose that there exists a subset $\mathcal{R}$ in  $\{1,\ldots k\}$ such that $\sum_{\ell\in\mathcal{R}}^\oplus \delta_{\ell}=0$. With a modified $k$-symbol combination that $\bar{\delta}_\ell=0$ for all $\ell\in\mathcal{R}$ and $\bar{\delta}_\ell=\delta_\ell$ for all $\ell\in\{1,\ldots k\}\setminus\mathcal{R}$, we have
\begin{equation}\label{eq:expurge}
{\sum_{\ell=1}^{k}}\tilde{\alpha}_{i,j_\ell}(\bar{\delta}_\ell)=\sum_{\ell\in\{1,\ldots k\}\setminus\mathcal{R}}\tilde{\alpha}_{i,j_\ell}({\delta}_\ell)\leq\sum_{\ell=1}^{k}\tilde{\alpha}_{i,j_\ell}({\delta}_\ell),
\end{equation}
where ${\sum^\oplus}_{\ell=1}^{k} \delta_\ell={\sum^\oplus}_{\ell=1}^{k} \bar{\delta}_\ell=\delta$. Thus the original combination can be ignored.
\end{Proof}
Directly following from Theorem \ref{thm:exclude-rep}, Lemma \ref{thm:max-order} shows that $\tilde{\beta}^{(k)}(\delta)$  of order $k>p$ is equal to $\tilde{\beta}^{(p)}(\delta)$, since the combinations with more than $p$ nonzero symbols can be ignored.
\begin{lemma}\label{thm:max-order}
With $q=2^p$, for all $\delta\in\text{GF}(q)$, we have
\begin{equation}\label{eq:high-order-equation2}\tilde{\beta}^{(p)}(\delta)=\tilde{\beta}^{(p+1)}(\delta)=\ldots=\tilde{\beta}(\delta)\end{equation}
\end{lemma}
\begin{Proof}
$\tilde{\beta}^{(k)}(\delta)$ is determined in (\ref{eq:norm-high-order}) by searching for the optimal $k$-symbol combination ${\sum^\oplus}_{\ell=1}^{k} \delta_\ell=\delta$. Assuming that some $\delta_\ell$ is $0$, this combination is equivalent to the $(k-1)$-symbol combination and has been considered for  $\tilde{\beta}^{(k-1)}(\delta)$. Otherwise if all symbols are nonzero, with $k\geq p+1$, we can consider the $p\times k$ binary matrix $\mathbf{B}$ of which the $\ell$-th column is the binary vector of $\delta_\ell$. Since the rank is at most $p$, it can be proved that  there must exist a subset $\mathcal{R}$ in  $\{1,\ldots k\}$ such that $\sum_{\ell\in\mathcal{R}}^\oplus \delta_{\ell}=0$. Following from Theorem \ref{thm:exclude-rep}, the $k$-symbol combination can be ignored, but the equivalent $(k-|\mathcal{R}|)$-symbol combination has been considered for $\tilde{\beta}^{(k-|\mathcal{R}|)}(\delta)$. Consequently, after ignoring every combination of more than $p$ nonzero symbols, the search space for $\tilde{\beta}^{(k)}(\delta)$ becomes equivalent to that for $\tilde{\beta}^{(p)}(\delta)$. It implies that $\tilde{\beta}^{(k)}(\delta)$ must be equal to $\tilde{\beta}^{(p)}(\delta)$.
\end{Proof}

By the derivations given above, we have proposed to reduce the
search space significantly in the deviation space, especially for
the larger check node degree and smaller field. Lemma
\ref{thm:max-order} also yields the maximal configuration order
required by (\ref{eq:orig-min-sum}), i.e., $\min(d_c-1,p)$.
Moreover, in (\ref{eq:norm-high-order}), the $k$ VNs are chosen from
$N_i\setminus j$ \emph{without} repetition. However, if $k$ VNs are
allowed to be chosen \emph{with} repetition, the search space will
expand such that (\ref{eq:norm-high-order}) can be approximated by
the lower bound:
\begin{flalign}
\tilde{\beta}^{(k)}(\delta)
&\geq\min_{{\sum^\oplus}_{\ell=1}^{k} \delta_\ell=\delta}\min_{j_1\in N_i\setminus j}\cdots\min_{j_k\in N_i\setminus j}\sum_{\ell=1}^{k}\tilde{\alpha}_{i,j_\ell}(\delta_\ell)\nonumber\\
&=\min_{{\sum^\oplus}_{\ell=1}^{k} \delta_\ell=\delta}\sum_{\ell=1}^{k}\min_{j_\ell \in N_i\setminus j}\tilde{\alpha}_{i,{j_\ell}}(\delta_\ell)\nonumber\\
&=\min_{{\sum^\oplus}_{\ell=1}^{k} \delta_\ell=\delta}\sum_{\ell=1}^{k}\tilde{\beta}^{(1)}(\delta_\ell), \label{eq:norm-high-order-approx-1}
\end{flalign}
where the last equation follows from (\ref{eq:chk-message-ord1}). Therefore, the SMSA can be carried out as follows:

\begin{Algorithm}\label{algorithm:smsa}
\noindent\textbf{The Simplified Min-Sum Algorithm}\\
\emph{Initialization}: Set $z_j = \arg\min_{d\in
\text{GF}(q)}{\lambda}_j(d)$. For all $i,j$ with $h_{i,j}\neq 0$, set ${a}_{i,j}=h_{i,j}\otimes z_j$ and $\tilde{\alpha}_{i,j}(h_{i,j}\otimes \delta) =
{\lambda}_j(\delta\oplus z_j)$. Set $\kappa=0$.
    \begin{itemize}
    \item Step 1) and 2) (The same as Step 1 and 2 in the EMSA)
    \end{itemize}
    \emph{CN processing: Step 3.1-4}
    \begin{itemize}
    \item Step 3.1) Compute the C2V hard messages:
     \begin{equation}\label{chk-proc-hd-smsa}{b}_{i,j}={\sum_{j'\in N_i\setminus j}}^\oplus{a}_{i,j'}.\end{equation}
     \item Step 3.2) Compute the step-$1$ soft messages:
    \begin{equation}\label{eq:chk-ord1}\tilde{\beta}^{(1)}_{i,j}(\delta)=\min_{j'\in N_i\setminus j}\tilde{\alpha}_{i,j'}(\delta).\end{equation}
    \item  Step 3.3) Compute the step-$2$ soft messages by selecting the combination of $k$ symbols according to Theorem \ref{thm:exclude-rep}:
    \begin{equation}\label{eq:chk-high-ord}\tilde{\beta}''_{i,j}(\delta)=\min_{{\sum^\oplus}_{\ell=1}^{k} \delta_\ell=\delta}\sum_{\ell=1}^{k}\tilde{\beta}^{(1)}_{i,j}(\delta_\ell).\end{equation}
    \item Step 3.4) Scaling and reordering: With $0< c\leq 1$, $\tilde{\beta}_{i,j}(\delta)\approx c\cdot\tilde{\beta}''_{i,j}(\delta)$.\\
    For $d\neq {b}_{i,j}$, ${\beta}_{i,j}(d)=\tilde{\beta}_{i,j}({b}_{i,j} \oplus d)$; otherwise ${\beta}_{i,j}({b}_{i,j})=0$.
    \item Step 4) (The same as Step 4 in the EMSA)
    \item Step 5) \emph{Message normalization and reordering:}
\begin{equation}{a}_{i,j}=\arg\min_{d\in\text{GF}(q)}\hat{\alpha}_{i,j}(d).\end{equation}
    \begin{equation}\label{eq:var-norm-sms}{\alpha}_{i,j}(d)=\hat{\alpha}_{i,j}(d)-\hat{\alpha}_{i,j}({a}_{i,j}).\end{equation}
     \begin{equation}\label{eq:var-reoder}\tilde{\alpha}_{i,j}(\delta)={\alpha}_{i,j}(\delta\oplus {a}_{i,j}).\end{equation}
    \item Step 6) (The same as the Step 6 in the EMSA)
    \item Go to Step 1.
    \end{itemize}
\end{Algorithm}

 As a result, the soft message generation is conducted in two steps (Step 3.2 and 3.3). To compute C2V messages $\tilde{\beta}_{i,j}$, first in Step 3.2 we compute the minimal entry values $\min_{j'} \tilde{\alpha}_{i,j'}(\delta)$ over all $j'\in N_i\setminus j$ for each $\delta\in \text{GF}(q)\setminus 0$. Then in Step 3.3, the minimal values are used to generate the approximation of $\tilde{\beta}_{i,j}(\delta)$. Instead of the configurations of all $d_c$ VNs in $N_i$, (\ref{eq:chk-high-ord}) optimizes over the combinations of $k$ symbols chosen from the field. Comparing Theorem \ref{lm:ord-k} to (\ref{eq:chk-ord1}) and (\ref{eq:chk-high-ord}), we can find that by our approximation method, in the SMSA, the optimization is performed over the VN set and symbol combination set separately and thus has the advantage of a much smaller search space.

\subsection{Practical Realization}\label{sec:realization}
 Because of the complexity issue, the authors of \cite{declercq2007decoding} suggested to use $k=4$ for (\ref{eq:orig-min-sum}), as using $k>4$ is reported to give unnoticeable performance improvement. Correspondingly,
 we only consider a small $k$ for (\ref{eq:chk-high-ord}). But it is still costly to generate all combinations with the large finite field. For example, with a 64-ary code there are totally $\displaystyle{q\choose 2}=2016$ combinations for $k=2$ and $\displaystyle{q\choose 4}=635376$ for $k=4$. Even with Theorem \ref{thm:exclude-rep} applied, the number of required combinations can be proved to be of $O(q^k)$.
 For this reason, we consider a reduced-complexity realization other than directly transforming the algorithm into the implementation. It can be shown that for $\delta'_1\oplus \delta'_2=\delta$  with $\delta'_1={\sum^\oplus}_{\ell=1}^{h} \delta_\ell$ and $\delta'_2={\sum^\oplus}_{\ell=h+1}^{k} \delta_\ell$ and $1< h < k$, in SMSA $\tilde{\beta}''(\delta)$ can also be approximated by
\begin{equation}\begin{split}
\tilde{\beta}''(\delta)
&\geq\min_{{\sum^\oplus}_{\ell=1}^{k} \delta_\ell=\delta}\left(\min_{\begin{subarray}{c}j_1,\ldots,j_{h}\in N_i\setminus j\\ j_1\neq\ldots\neq j_{h} \end{subarray}}\sum_{\ell=1}^{h}\tilde{\alpha}_{i,{j_\ell}}(\delta_\ell)+\min_{\begin{subarray}{c}j_{h+1},\ldots,j_{k}\in N_i\setminus j\\ j_{h+1}\neq\ldots\neq j_{k} \end{subarray}}\sum_{\ell=h+1}^{k}\tilde{\alpha}_{i,{j_\ell}}(\delta_\ell)\right)\\
&=\min_{\delta'_1 \oplus \delta'_2=\delta}\left(\min_{{\sum^\oplus}_{\ell=1}^{h} \delta_\ell=\delta'_1}\min_{\begin{subarray}{c}j_1,\ldots,j_{h}\in N_i\setminus j\\ j_1\neq\ldots\neq j_{h} \end{subarray}}\sum_{\ell=1}^{h}\tilde{\alpha}_{i,{j_\ell}}(\delta_\ell)+\min_{{\sum^\oplus}_{\ell=h+1}^{k} \delta_\ell=\delta'_2}\min_{\begin{subarray}{c}j_{h+1},\ldots,j_{k}\in N_i\setminus j\\ j_{h+1}\neq\ldots\neq j_{k} \end{subarray}}\sum_{\ell=h+1}^{k}\tilde{\alpha}_{i,{j_\ell}}(\delta_\ell)\right)\\
&\geq\min_{\delta'_1 \oplus \delta'_2=\delta}\left(\tilde{\beta}'(\delta'_1)+\tilde{\beta}'(\delta'_2)\right),
\label{eq:norm-high-order-approx-2}\end{split}\end{equation}
where $\tilde{\beta}'(\delta)$ denotes the primitive message, that is the soft message of any order lower than the required order $k$. Hence we can successively combine two 2-symbol combinations to make a 4-symbol one by two sub-steps with a look-up table (LUT), in which all 2-symbol combinations are listed. This method allows us to obtain $k$-symbol combinations using $\log_2 k$ sub-steps, with $k$ equal to a power of $2$. Based on this general technique, in the following we will select $k$ to meet requirements for complexity and performance, and then practical realizations are provided specifically for different $k$.

The approximation loss with a small $k$ results from the reduced search, with the search space size of $O(q^k)$. According to Theorem \ref{thm:max-order}, the full-size search space is of $p$-symbol combinations, with the size of $O(q^p)$. As the size ratio between two spaces is of $O(q^{p-k})$, the performance degradation is supposed to be smaller for smaller fields. $k=1$ was shown to have huge performance loss for NB-LDPC codes \cite{declercq2007decoding}. By the simulation results in Section \ref{sec:simulation}, setting $k=2$ will be shown to have smaller loss with smaller fields when compared to the EMSA. And having $k=4$ will be shown to provide negligible loss, with field size $q$ up to $256$. Since we observed that using $k>4$ gives little advantage, two settings $k=2$ and $k=4$ will be further investigated in the following as two tradeoffs between complexity and performance.

\begin{table}[t]
\caption{The look-up table $D$ for GF($2^3$).} \label{tbl:gf8:lut}
\centering
\begin{tabular}{c|cc|cc}
  $\delta \backslash f $ & 0 & 1 & 2 & 3 \\
  \hline
  1 & (0,1) & (2,3) & (4,5) & (6,7) \\
  2 & (0,2) & (1,3) & (4,6) & (5,7) \\
  3 & (0,3) & (1,2) & (4,7) & (5,6) \\
  \hline
  4 & (0,4) & (1,5) & (2,6) & (3,7) \\
  5 & (0,5) & (1,4) & (2,7) & (3,6) \\
  6 & (0,6) & (1,7) & (2,4) & (3,5) \\
  7 & (0,7) & (1,6) & (2,5) & (3,4) \\
  \hline
\end{tabular}
\end{table}
\begin{algorithm}[t]
\caption{Generate the look-up table for GF($q$).} \label{alg:lut}
{\small
\begin{algorithmic}[1]
  \FOR{$\delta'=1\ldots q-1$}
  \FOR{$\delta''=(\delta'\oplus 1)\ldots q-1$}
  \STATE $\delta=\delta'\oplus \delta''$;
  \STATE $D(\delta).\text{Add}(\delta',\delta'')$;
  \ENDFOR
  \ENDFOR
\end{algorithmic}}
\end{algorithm}

Let us first look at the required LUT. Shown in Algorithm \ref{alg:lut}, the pseudo code generates the list of combinations $(\delta_1, \delta_2)$ without repetition for each target $\delta$ with $\delta_1\oplus\delta_2=\delta$. Since we have $q/2$ combinations for each of  $q-1$ target, $D$ can be depicted as a two-dimensional table with $q-1$
rows and $q/2$ columns. For $1\leq d\leq (q-1)$ and $0\leq f\leq
q/2-1$, each cell $D_{\delta,f}$ in the table is a two-tuple containing
two elements $D_{\delta,f}(0)$ and $D_{\delta,f}(1)$, which satisfy the addition rule
$D_{\delta,f}(0) \oplus D_{\delta,f}(1) = \delta$. For example, when $q=8$, the LUT is provided in
Table~\ref{tbl:gf8:lut}.

Step 3.3 and (\ref{eq:chk-high-ord}) can be realized by Step 3.3.1 and 3.3.2 given below.

\begin{itemize}
\item Step 3.3.1) With the LUT $D$, compute the step-$1$ messages by
\end{itemize}
\begin{equation}\label{eq:chk-high-ord-table1}
\tilde{\beta}'_{i,j}(\delta) =
\min_{f=0\ldots q/2-1} \Big(
\tilde{\beta}^{(1)}_{i,j}(D_{\delta,f}(0))+\tilde{\beta}^{(1)}_{i,j}(D_{\delta,f}(1))\Big).
\end{equation}
\begin{itemize}
\item Step 3.3.2) Compute the step-$2$ messages by
\end{itemize}
\begin{equation}\label{eq:chk-high-ord-table2}
\tilde{\beta}''_{i,j}(\delta)=
\min_{f=0\ldots q/4-1} \Big(
\tilde{\beta}'_{i,j}(D_{\delta,f}(0))+\tilde{\beta}'_{i,j}(D_{\delta,f}(1))\Big).
\end{equation}
By the definition, we let $\tilde{\beta}^{(1)}_{i,j}(0)=\tilde{\beta}'_{i,j}(0)=0$, so $\tilde{\beta}^{(1)}_{i,j}(D_{\delta,0}(0))+\tilde{\beta}^{(1)}_{i,j}(D_{\delta,0}(1))=\tilde{\beta}^{(1)}_{i,j}(\delta)$.

The first sub-step combines two symbols $D_{\delta,f}(0)$ and $D_{\delta,f}(1)$ for each $\delta$ and $f$, making a 2-symbol combination. The comparison will be conducted over $f=0\ldots q/2-1$ for each $\delta$. Assume that the index of the minimal value is  $f^*(\delta)$. Then the second sub-step essentially combines two two-tuples $ D_{D_{\delta,f}(0),f^*(D_{\delta,f}(0))}$ and $D_{D_{\delta,f}(1),f^*(D_{\delta,f}(1))}$, making a 4-symbol combination. It can be proved that all 4-symbol combinations can be considered by combining two-tuples $D_{\delta,f}$ of $f=0,1,\ldots q/4-1$. So the second sub-step only performs the left half of the Table $D$. For instance, over $\text{GF}(2^3)$ the left half of $D$ is formed by $f=0,1$ in Table~\ref{tbl:gf8:lut}.

For $k=2$ and $k=4$ respectively, we define two versions of SMSA, i.e., the one-step SMSA (denoted as SMSA-1) and the two-step SMSA (denoted as SMSA-2). The SMSA-1 is the same as the SMSA-2 except for the implementation of Step 3.3. The SMSA-1 only requires Step 3.3.1 and skips Step 3.3.2, while the SMSA-2 implements both steps. We will present the performance and complexity results of the SMSA-1 and SMSA-2 in the following sections.

\section{Simulation Results}

\label{sec:simulation}
In this section, we use five examples to demonstrate the
performance of the above proposed SMSA for decoding NB-LDPC codes.
The existing algorithms including the QSPA, EMSA, and MMA are used
for performance comparison.  The SMSA includes the one-step (SMSA-1) and two-step (SMSA-2) versions. In the first two examples, three codes over GF($2^4$), GF($2^6$), and GF($2^8$)
are considered. We show that
the SMSA-2  has very good performance for different finite fields and
modulations. And the SMSA-1 has small performance loss compared to the SMSA-2 over GF($2^4$) and GF($2^5$). The binary phase-shift keying
(BPSK) and quadrature amplitude modulation (QAM) are applied over
the additive white Gaussian noise (AWGN) channel.  In the third
example, we study the fixed-point realizations of SMSA and find that
it is exceptionally suitable for hardware implementation.  The fourth example compares
the performance of the SMSA, QSPA, EMSA, and MMA over the uncorrelated  Rayleigh-fading channel. The SMSA-2 shows its reliability with higher channel randomness. In the
last example, we research on the convergence speed of SMSA and show
that it converges almost as fast as EMSA.

\begin{figure}[h]
  \centering
  \includegraphics[width=0.7\textwidth]{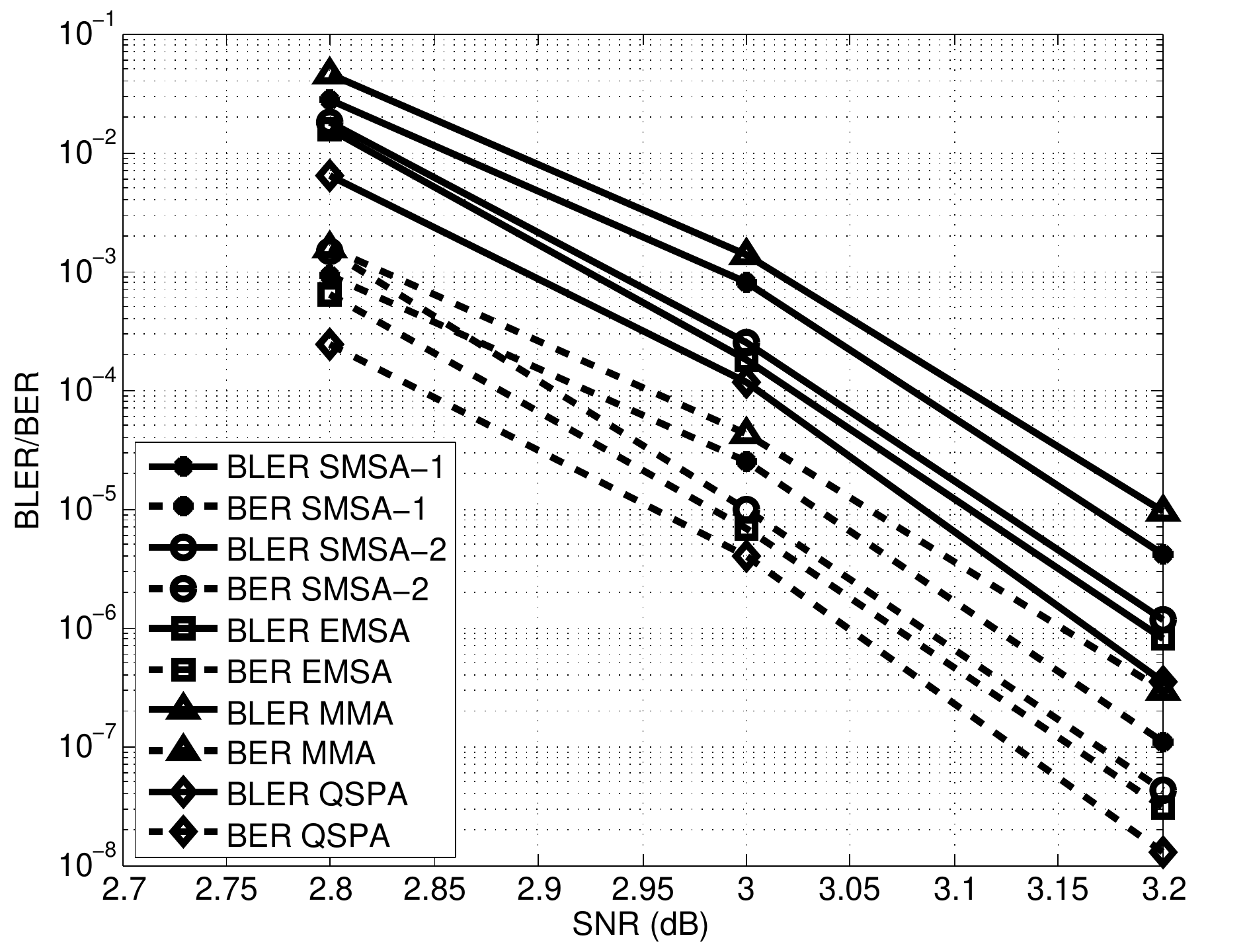}
  \caption{BLER and BER comparison of the SMSA-1, SMSA-2, EMSA, MMA, and QSPA with the (1057,813) code over $\text{GF}(2^4)$. The BPSK is used over the AWGN channel. The maximal iteration number $\kappa_{\text{max}}$ is set to 50.}
  \label{fig:gf16}
\end{figure}

\begin{figure}[h]
  \centering
  \includegraphics[width=0.7\textwidth]{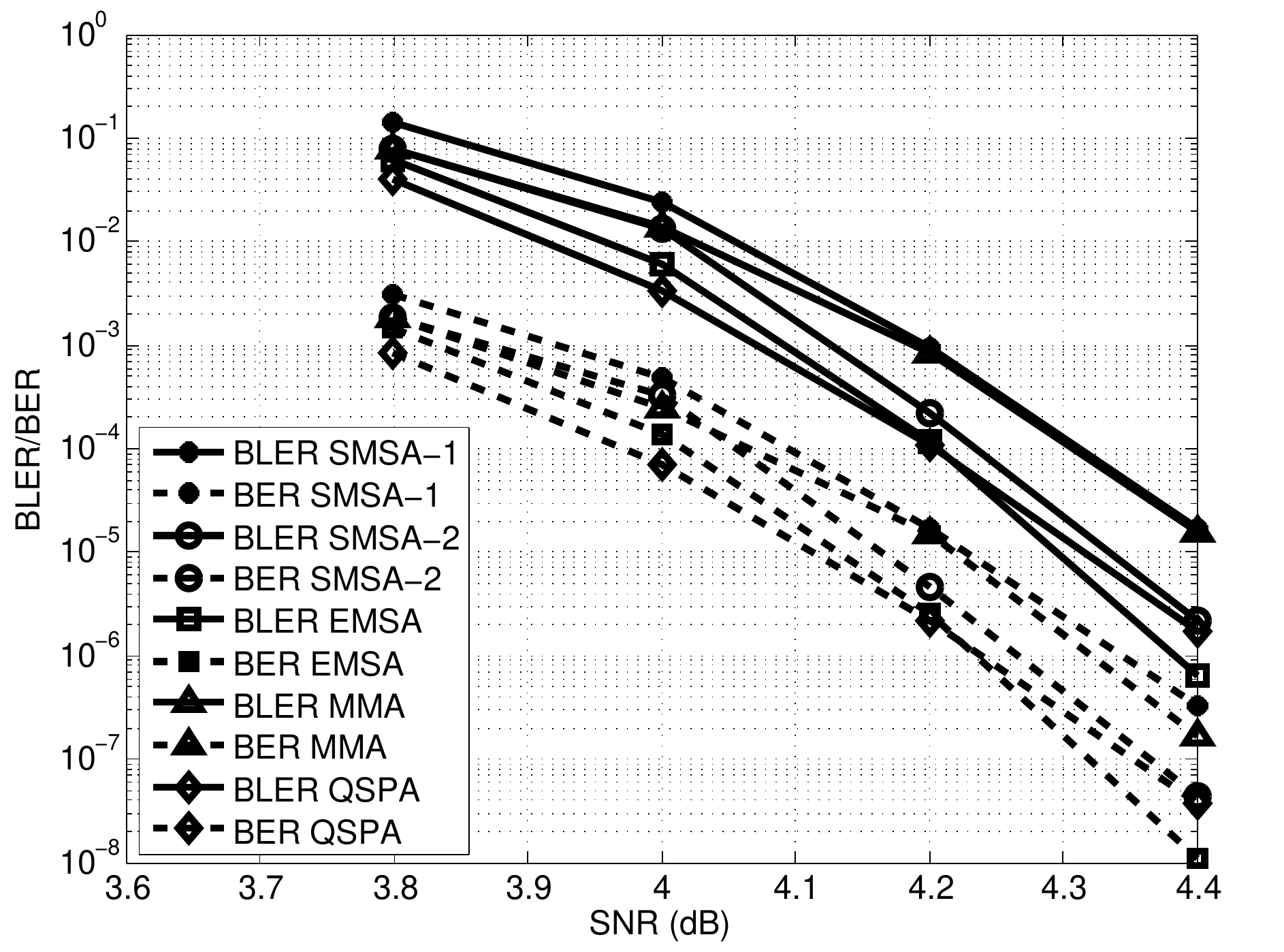}
  \caption{BLER and BER comparison of the SMSA-1, SMSA-2, EMSA, MMA, and QSPA with the (495,433) code over $\text{GF}(2^6)$. The BPSK is used over the AWGN channel. The maximal iteration number $\kappa_{\text{max}}$ is set to 50.}
  \label{fig:gf64}
\end{figure}

\begin{figure}[h]
  \centering
  \includegraphics[width=0.7\textwidth]{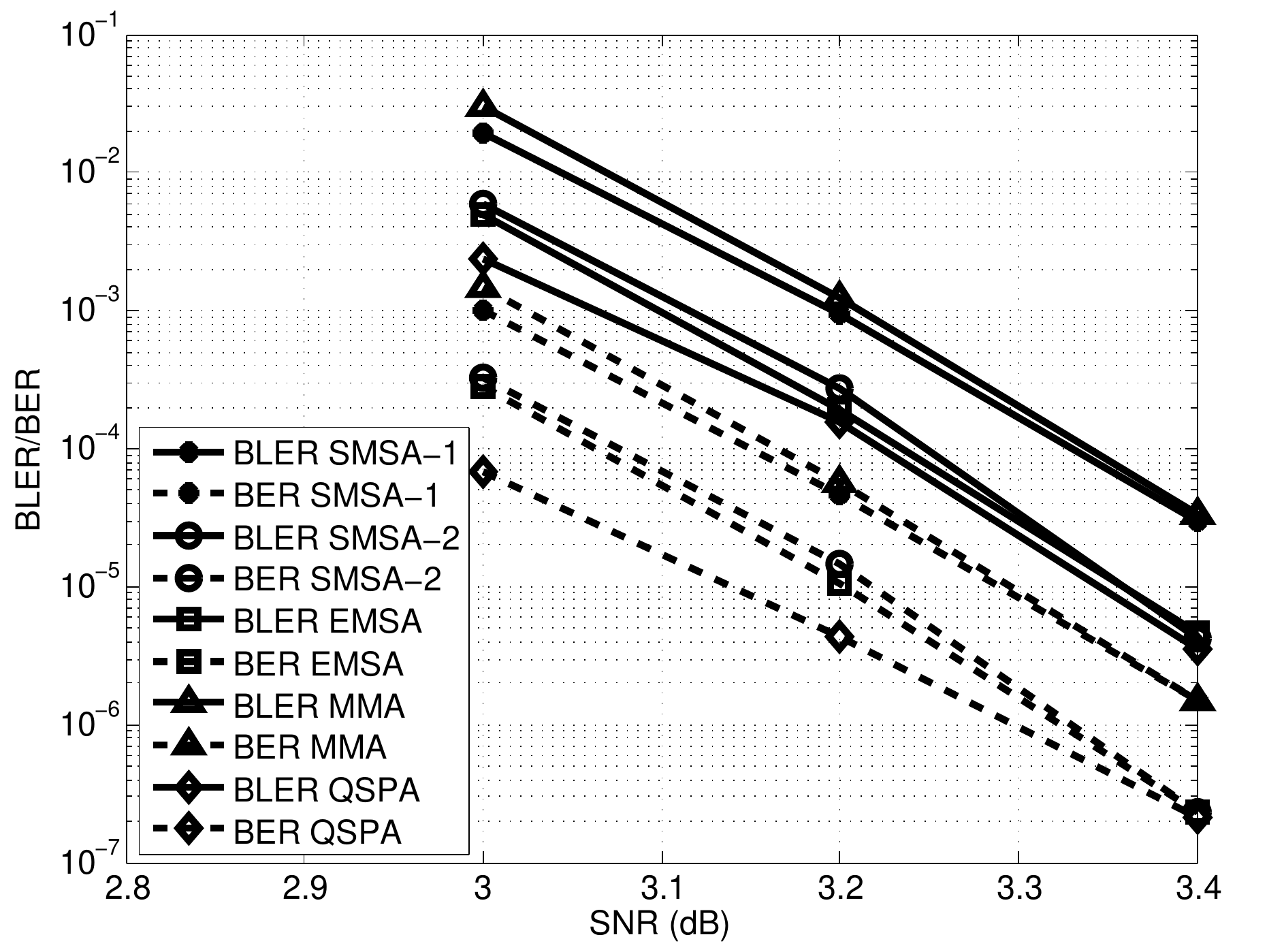}
  \caption{BLER and BER comparison of the SMSA-1, SMSA-2, EMSA, MMA, and QSPA with the (273,191) code over $\text{GF}(2^8)$. The BPSK is used over the AWGN channel. The maximal iteration number $\kappa_{\text{max}}$ is set to 50.}
  \label{fig:gf256}
\end{figure}

\begin{example}\label{ex:bpsk}
\emph{(BPSK-AWGN)} Three codes constructed by computer search over
different finite fields are used in this example.  Four
iterative decoding algorithms (SMSA, QSPA, EMSA, and MMA) are simulated with
the BPSK modulation over the binary-input AWGN channel for every
code.
The maximal iteration number $\kappa_{\text{max}}$ is set to
$50$ for all algorithms.  The bit error rate (BER) and
block error rate (BLER) are obtained to characterize the error
performance.  The first code is a rate-0.769 (3,13)-regular
(1057,813) code over $\text{GF}(2^4)$, and its error performance is
shown in Fig.~\ref{fig:gf16}.  We use optimal scaling factors
$c=0.60$, $0.75$, and $0.73$ for the SMSA-1, SMSA-2, and EMSA respectively.  The second
code is a rate-0.875 (3,24)-regular (495,433) code over
$\text{GF}(2^6)$, and its error performance is shown in
Fig.~\ref{fig:gf64}.  We use optimal scaling factors $c=0.50$, $0.70$, and
$0.65$ for the SMSA-1, SMSA-2, and EMSA respectively.  The third code is a
rate-0.70 (3,10)-regular (273,191) code over $\text{GF}(2^8)$, and
its error performance is shown in Fig.~\ref{fig:gf256}.  We use
optimal scaling factors $c=0.35$, $0.575$, and $0.60$ for the SMSA-1, SMSA-2, and EMSA
respectively. Taking the EMSA as a benchmark at BLER of $10^{-5}$,  we observe that the SMSA-2
has SNR loss of less than 0.05 dB, while the
MMA suffers from about 0.1 dB loss. The SMSA-1 has 0.06 dB loss with $\text{GF}(2^4)$ and almost 0.15 dB loss with $\text{GF}(2^6)$ and $\text{GF}(2^8)$ against the EMSA. As discussed in Section \ref{sec:realization}, the SMSA-1 performs better with smaller fields.
At last, the QSPA has SNR gain of less
than 0.05 dB and yet is viewed as undesirable for
implementation.
\end{example}

\begin{figure}[h]
  \centering
  \includegraphics[width=0.7\textwidth]{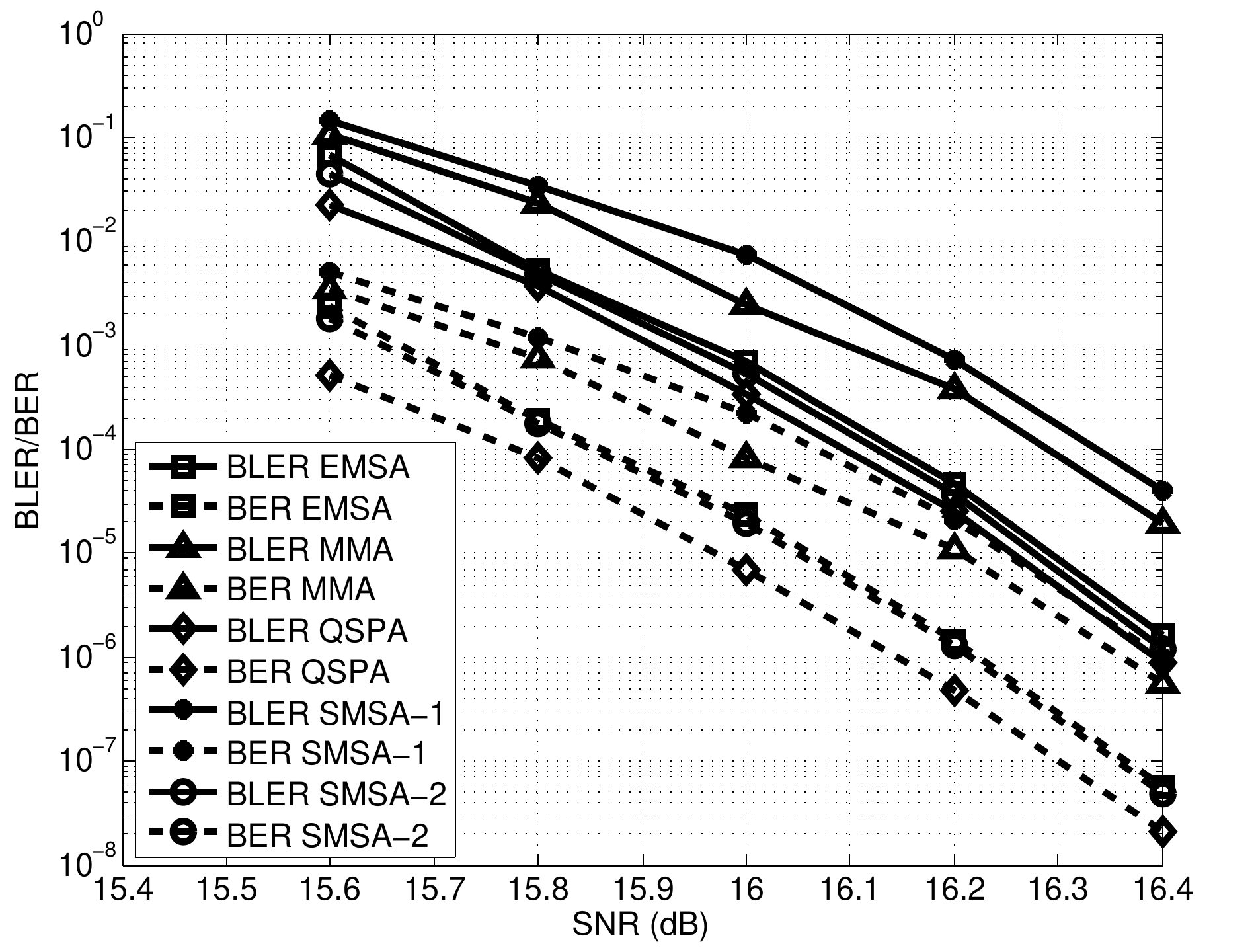}
  \caption{BLER and BER comparison of the SMSA-1, SMSA-2, EMSA, MMA, and QSPA with the (495,433) code over $\text{GF}(2^6)$. The 64-QAM is used over the AWGN channel. The maximal iteration number $\kappa_{\text{max}}$ is set to 50.}
  \label{fig:64qam}
\end{figure}

\begin{example}
\emph{(QAM-AWGN)} Fig.~\ref{fig:64qam} shows the performance of the
$64$-ary (495,433) code, the second code in
Example~\ref{ex:bpsk}, with the rectangular 64-QAM.  Four decoding algorithms (SMSA, QSPA, EMSA,
and MMA) are simulated with finite field
symbols directly mapped to the grey-coded constellation symbols
over the AWGN channel.  The maximal iteration number
$\kappa_{\text{max}}$ is set to $50$ for all algorithms.  The SMSA-1, SMSA-2, and EMSA have the optimal scaling factors $c=0.37$, $0.60$, and $0.50$ respectively.We
note that the SMSA-2 and EMSA achieve nearly the same BER and BLER,
while the MMA and SMSA-1 have 0.11 and 0.14 dB of performance loss.
\end{example}

\begin{figure}[h]
\centering
\includegraphics[width=0.70\textwidth]{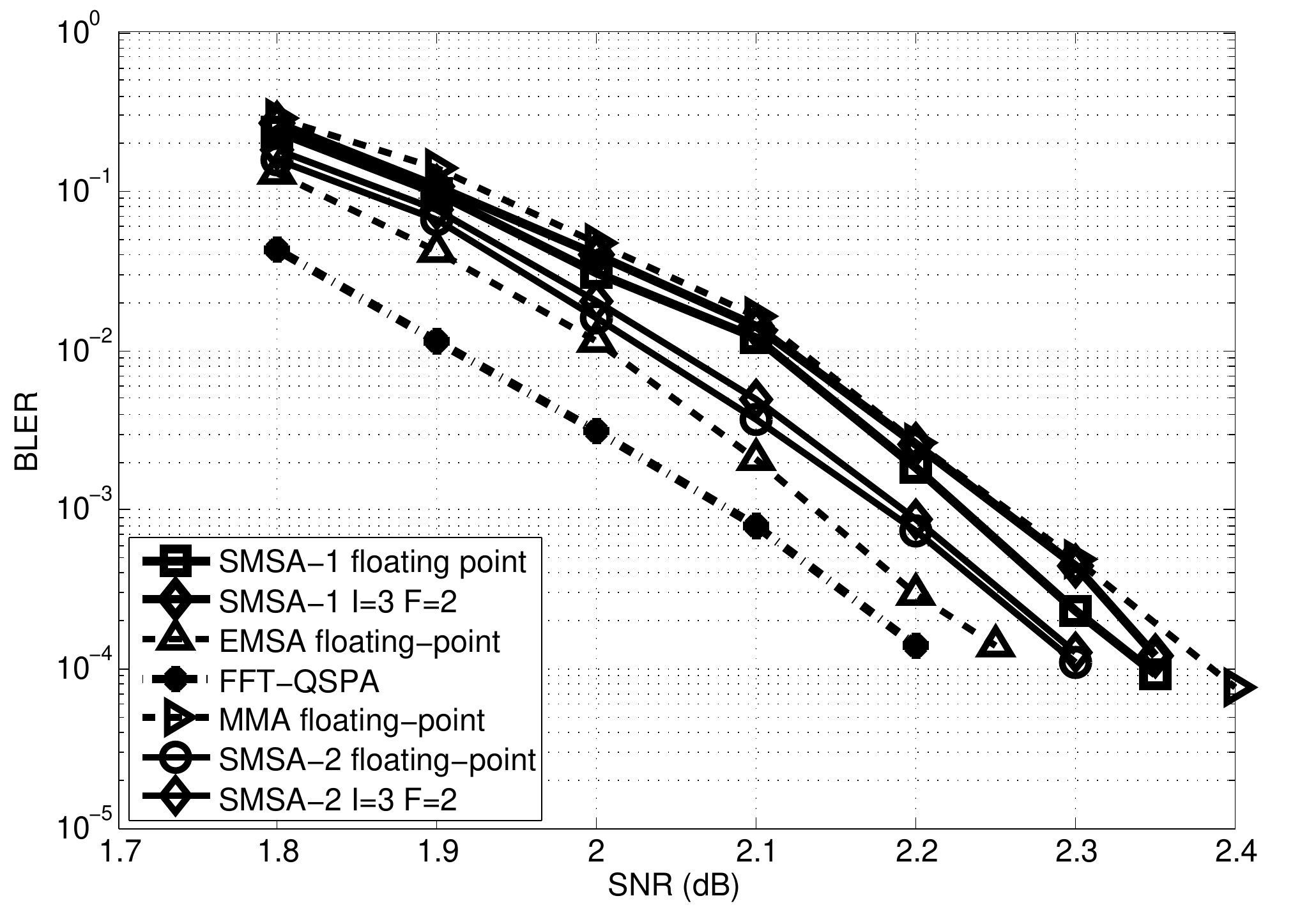}
\caption{BLER comparison of the SMSA-1, SMSA-2 (fixed-point and
floating-point), QSPA, EMSA, and MMA (floating-point only) with the
(620,310) code over GF($2^5$). The BPSK is used over the AWGN channel. The maximal
iteration number $\kappa_{\text{max}}$ is set to 50.}
\label{fig:620:310}
\end{figure}

\begin{example}\label{ex:620:310}\emph{(Fixed-Point Analysis)}
To investigate the effectiveness of the SMSA, we evaluate the block
error performance of the (620,310) code over $\text{GF}(2^5)$ taken
from \cite{zhou2009construction}.  The parity-check matrix of the
code is a $10\times 20$ array of $31\times 31$ circulant permutation
matrices and zero matrices. The floating-point QSPA, EMSA, MMA, SMSA-1, and SMSA-2 and the fixed-point SMSA-1 and SMSA-2 are simulated using the BPSK
modulation over the AWGN channel.
The BLER results are shown in
Fig.~\ref{fig:620:310}.  The optimal scaling factors for
the SMSA-1, SMSA-2, and EMSA are $c=0.6875$, $0.6875$, and $0.65$ respectively. The maximal iteration number
$\kappa_{\text{max}}$ is set to $50$ for all algorithms. Let $I$ and $F$ denote the number of
bits for the integer part and fraction part of the quantization
scheme. We observe that for SMSA-1 and SMSA-2 \emph{five} bits ($I=3$, $F=2$) are
sufficient. For approximating the QSPA and EMSA, the SMSA-2 has SNR loss of only 0.1 dB and
0.04 dB  at BLER of $10^{-4}$, respectively.
And the SMSA-1 has SNR loss of 0.14 dB and 0.08 dB respectively.
\end{example}

\begin{figure}[h]
  \centering
  \includegraphics[width=0.7\textwidth]{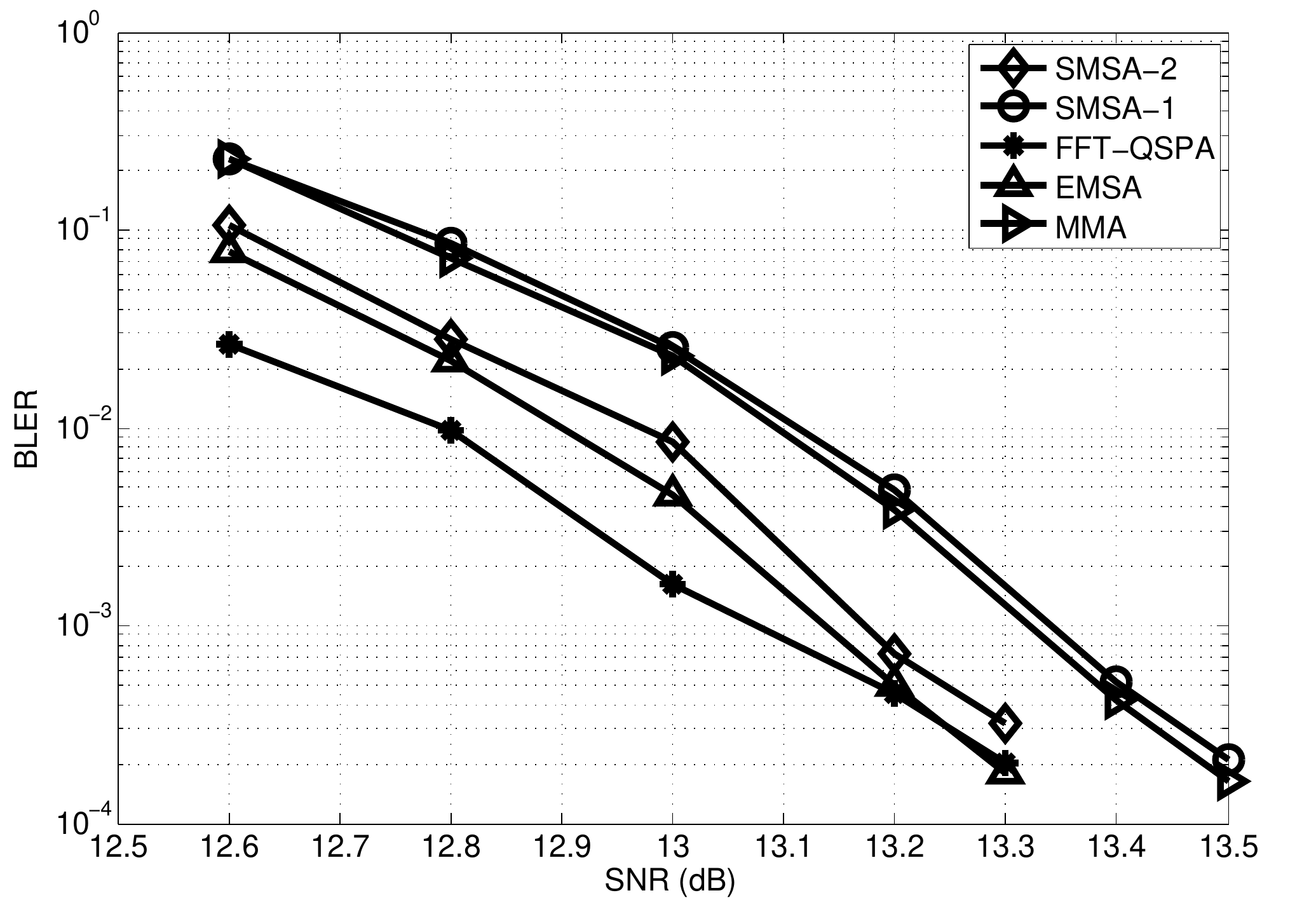}
  \caption{BLER comparison of the SMSA-1, SMSA-2, QSPA, EMSA, and MMA with the
(620,310) code over GF($2^5$). The BPSK is used over the uncorrelated Rayleigh-fading channel. The maximal
iteration number $\kappa_{\text{max}}$ is set to 50.}
  \label{fig:620:310-fading}
\end{figure}

\begin{example}\label{ex:620:310}\emph{(Fading Channel)}
To test the reliability of the SMSA, we examine the error performance
of the $32$-ary (620,310) code given
in Example 3 over the uncorrelated Rayleigh-fading channel with additive Gaussian noise. The channel information is assumed to be known to the receiver. The floating-point QSPA, EMSA, MMA,
SMSA-1, and SMSA-2 are simulated using the BPSK
modulation, as the BLER results are shown in
Fig.~\ref{fig:620:310-fading}. Compared to the EMSA, the SNR loss of SMSA-2 is within 0.1 dB, while the SMSA-1 and MMA have around 0.2 dB loss.
The QSPA has performance gain in low and medium SNR regions and no gain at high SNR.
\end{example}

\begin{figure}[h]
  \centering
  \includegraphics[width=0.7\textwidth]{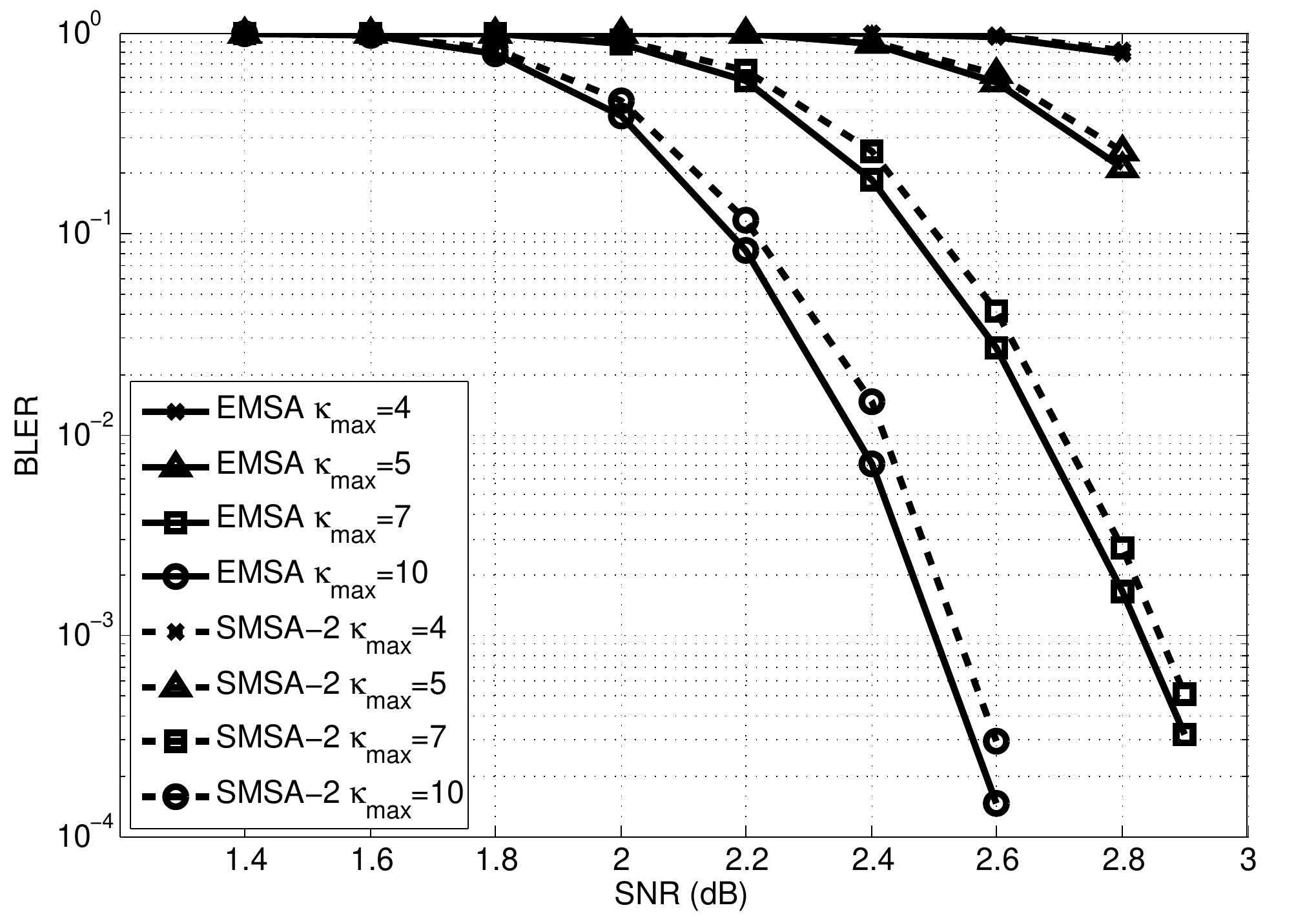}
  \caption{BLER comparison of the SMSA-2 and EMSA with the (620,310) code over
GF($2^5$). The BPSK is used over the AWGN channel. The maximal iteration number $\kappa_{\text{max}}$ is set to 4, 5, 7, and 10.}
  \label{fig:620:310-limitediter}
\end{figure}

\begin{figure}[h]
  \centering
  \includegraphics[width=0.7\textwidth]{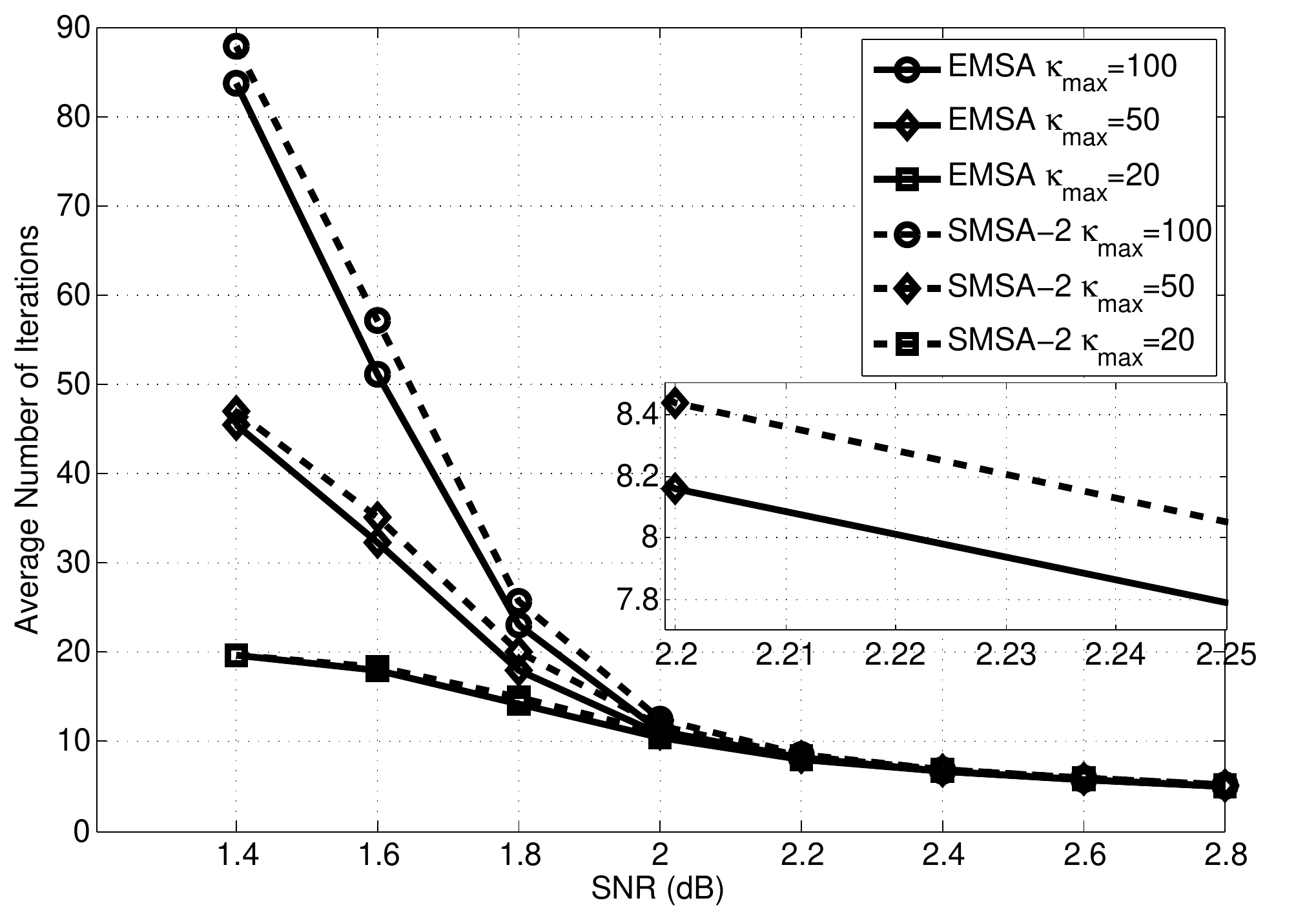}
  \caption{The average number of iterations for the SMSA-2 and EMSA with the (620,310) code over
GF($2^5$). The BPSK is used over the AWGN channel. The maximal iteration number
$\kappa_{\text{max}}$ is set to 20, 50, and 100.}
  \label{fig:620:310-avgiter}
\end{figure}

\begin{example}\label{ex:limitediter}
\emph{(Convergence Speed)} Consider again the 32-ary (620,310) code given
in Example 3. The block error performances for this code using the
SMSA-2 and EMSA with 4, 5, 7, and 10 maximal iterations are shown in
Fig.~\ref{fig:620:310-limitediter}.  At BLER of
$10^{-3}$, the SNR gap between the SMSA-2 and EMSA is 0.04
dB for various $\kappa_{\text{max}}$.
To further investigate the
convergence speed, we summarize the average number of iterations for
the EMSA and SMSA-2 with 20,  50, and 100 maximal iterations and show the
results in Fig.~\ref{fig:620:310-avgiter}.  It should be noted that shown in Fig.~\ref{fig:620:310},
the SNR gap of BLER between EMSA and SMSA-2 is about 0.04 dB, at BLER of $10^{-3}$ and SNR of about 2.2 dB . By examining the curves of Fig.~\ref{fig:620:310-avgiter} at SNR of about 2.2 dB (in the partial enlargement), we observe that for the same average iteration number the difference of required SNR is also around 0.04 dB between the two algorithms. Since a decoding failure increases the average iteration number, the SNR gap of error performance can be seen as the main reason for the SNR gap of average iteration numbers. Therefore, as the failure occurs often at low SNR and rarely at high SNR, in Fig.~\ref{fig:620:310-avgiter} the iteration increase for SMSA-2 at high SNR is negligible ($<5\%$ at 2.2 dB), and at low and medium SNR the gap is larger ($\approx 11\%$ at 1.8 dB). Although the result is not shown, we observe that the SMSA-1 also has similar convergence properties, and the iteration increase compared with the EMSA at high SNR is around $6\%$.
\end{example}

\section{Complexity Analysis}\label{sec:complexity}
In this section, we analyze the computational complexity of the SMSA
and compare it with the EMSA.
The comparison of average required iterations is provided in Example \ref{ex:limitediter} of Section \ref{sec:simulation}.
With a fixed SNR, the SMSA requires slightly more ($5\sim 6\%$) number of average iterations than the EMSA at medium and high SNR region. As the two algorithms have small (within 0.2 dB) performance difference, especially between the EMSA and SMSA-2, we think that it is fair to simply compare the complexity of the SMSA and EMSA by the computations per iteration. Moreover, since the VN processing is similar
for both algorithms, we only analyze the CN processing.  The
required operation counts per iteration for a CN with degree $d_c$ are
adopted as the metric.

To further reduce the duplication of computations in CN processing,
we propose to transform the Step 3.1 and 3.2 of SMSA as follows.
Step 3.1 can be transformed into two sub-steps.  We define $${A}_{i}
= {\sum_{j'\in N_i}}^{\oplus} a_{i,j'}.$$  Then each ${b}_{i,j}$ can
be computed by $${b}_{i,j} = a_{i,j} {\oplus} A_i.$$  Thus totally
it takes $2d_c - 1$ finite field additions to compute this step for
a CN.

Similarly, the computation of Step 3.2 can be transformed into two
sub-steps. For the $i$-th row of the parity-check matrix, we define
a three-tuple
$\{\text{min1}_i(\delta),\text{min2}_i(\delta),\text{idx}_i(\delta)\}$,
in which $$\text{min1}_i(\delta) \equiv \min_{j'\in
N_i}\tilde{\alpha}_{i,j'}(\delta),$$
$$\tilde{\alpha}_{i,\text{idx}_i(\delta)}(\delta) \equiv
\text{min1}_i(\delta),$$ $$\text{min2}_i(\delta) \equiv \min_{j'\in
N_i\setminus \text{idx}_i(\delta)}\tilde{\alpha}_{i,j'}(\delta).$$
For each nonzero symbol $\delta$ in $\text{GF}(q)$, it takes at most
$1+2(d_c-2)=2d_c-3$ min operations, and each operation can be
realized by a comparator and multiplexor to compute the 3-tuple
$\{\text{min1}_i(\delta),\text{min2}_i(\delta),\text{idx}_i(\delta)\}.$

The remaining computations of Step 3.2 can be computed equivalently
by $$\tilde{\beta}^{(1)}_{i,j}(\delta) = \text{min1}_i(\delta) \quad
\text{if } j\neq \text{idx}_i(\delta);$$
$$\tilde{\beta}^{(1)}_{i,j}(\delta) = \text{min2}_i(\delta) \quad
\text{if } j= \text{idx}_i.$$ It takes $d_c$ comparisons and $d_c$
two-to-one selections to perform the required operations. As there
are $q-1$ entries of $\tilde{\beta}^{(1)}_{i,j}$, the overall
computations of Step 3.2 per CN requires $(3d_c-3)(q-1)$ comparators
and $(3d_c-3)(q-1)$ multiplexors.

For the SMSA-2, Step 3.3 is realized by Step 3.3.1 and 3.3.2. To compute Step 3.3.1
for each symbol $\delta$, it takes $(q-2)/2$ summations and
$(q-2)/2$ comparisons. To compute Step 3.3.2 for each $\delta$, it
takes $(q-4)/4$ summations and $(q-4)/4$ comparisons. Therefore,
totally it takes $3q/4-2$ summations and min operations for each
$\delta$. As we have $q-1$ nonzero symbols in $\text{GF}(q)$,
overall it requires $(3q/4-2)(q-1)d_c$ summations, comparisons, and two-to-one
selections. For the SMSA-1, it requires $(q/2-1)(q-1)d_c$ summations, comparisons, and two-to-one
selections. Step 3.4 performs scaling and shifting and thus is
ignored here, since the workload is negligible compared to LLR
calculations.

\begin{table*}[t]
\caption{The required operations per iteration and memory usage to perform the CN
processing of a CN with degree $d_c$ for a $q$-ary
code. The bit width per sub-message is $w$.}
\label{tbl:complexity} \centering \small
\begin{tabular}{c|c|c|c}
\hline
 Type & SMSA-1 & SMSA-2 & EMSA \\
\hline
Finite Field Additions & $2d_c - 1$ & $2d_c - 1$ & 0 \\
Summations & $(q/2-1)(q-1)d_c$ & $(3q/4-2)(q-1)d_c$ & $3(d_c-2)q^2$ \\
Comparisons and Selections & $((q/2+2)d_c-3)(q-1)$ & $((3q/4+1)d_c-3)(q-1)$  & $3(d_c-2)q(q-1)$ \\
Memory Usage (Bits) & $(2w+\lceil\log_2 d_c\rceil)(q-1)$ & $(2w+\lceil\log_2 d_c\rceil)(q-1)$ & $w d_c q$ \\
\; & $+p d_c$ & $+p d_c$ & \; \\
\hline
\end{tabular}
\end{table*}

Then let us analyze the CN processing in EMSA for comparison. As in
\cite{wang:tcas1,zhang:tcas1,Xiaoheng:TCASI:Nonbinary}, usually the
forward-backward scheme is used to reduce the implementation
complexity. For a CN with degree $d_c$, $3(d_c-2)$ stages are
required, and each stage needs $q^2$ summations and $(q-1)q$ min
operations. Overall, in the EMSA, each CN has $q^2 d_c$ summations,
$(q-1)q d_c$ comparisons, and $(q-1)q d_c$ two-to-one selections.
The results for the SMSA and EMSA are summarized in
Table~\ref{tbl:complexity}.  As in implementation the required
finite field additions of SMSA take only marginal area, we see that
the SMSA requires much less computations compared to the EMSA.

Since the computational complexity for decoding NB-LDPC codes is
very large, the decoder implementations usually adopt
partially-parallel
architectures.
Therefore, the CN-to-VN messages are usually stored in the decoder
memory for future VN processing.  As memory occupies significant
amount of silicon area in hardware
implementation,
optimizing the memory usage becomes an important research problem~\cite{wang:tcas1,zhang:tcas1,Xiaoheng:TCASI:Nonbinary}.
For Step 3.2 of SMSA, the 3-tuple
$\{\text{min1}_i(\delta),\text{min2}_i(\delta),\text{idx}_i(\delta)\}$
can be used to recover the messages
$\tilde{\beta}^{(1)}_{i,j}(\delta)$ for all $j\in N_i$.  Assume that the bit width for
each entry of the soft message is $w$ in the CN processing. Then for
each $\delta$ in $\text{GF}(q)$, the SMSA needs to store
$2w+\lceil\log_2 d_c\rceil$ bits for the 3-tuple. Also, it needs to
store the hard messages ${a}_{i,j}$ in Step 3.1, which translate to
$p\times d_c$ bits of storage. To store the intermediate messages
for the CN processing of each row, totally the SMSA requires to
store $(2w+\lceil\log_2 d_c\rceil)(q-1)+p d_c$ bits. In comparison,
for the EMSA, there is no correlation between ${\beta}_{i,j}(d)$ of
each $j\in N_i$ in the $i$-th CN. Therefore, the EMSA requires to
store the soft messages ${\alpha}_{i,j}(d)$ of all $j\in N_i$, which
translate to $w\times d_c \times q$ bits.  We see that the SMSA
requires much less memory storage compared to the EMSA.

We take as an example the (620,310) code over $\text{GF}(2^5)$ used in Section \ref{sec:simulation}.
With $w=5$ and $d_c=6$, the SMSA-1 requires 2790 summations, 3255 comparisons, and 433 memory bits for each CN per iteration,
and the SMSA-2 requires 4092 summations, 4557 comparisons, and 433 memory bits. The EMSA requires 12288 summations, 11904 comparisons, and 960
memory bits. As a result, compared to the EMSA, the SMSA-1 saves 77$\%$ on summations and 73$\%$ on comparisons, and the SMSA-2 saves 67$\%$ and 62$\%$ respectively. Both of the two SMSA versions save
55$\%$ on memory bits. More hardware implementation results are presented for SMSA-2 in \cite{xhchen:tcas1:smsa}, which shows exceptional saving in silicon area when compared with existing NB-LDPC decoders.

\section{Conclusions} \label{sec:conclusion}

In this paper, we have presented a hardware-efficient decoding
algorithm, called the SMSA, to decode NB-LDPC codes. This algorithm
is devised based on significantly reducing the search space of
combinatorial optimization in the CN processing.  Two practical realizations, the one-step and two-step SMSAs,
are proposed for effective complexity-performance tradeoffs. Simulation results
show that with field size up to 256, the two-step SMSA has negligible error performance loss compared to
the EMSA over the AWGN and Rayleigh-fading channels. The one-step SMSA has 0.1 to 0.2 dB loss depending on the field size. Also, the
fixed-point study and convergence speed research show that it is
suitable for hardware implementation.  Another important feature of
SMSA is simplicity.  Based on our analysis, the SMSA has much lower
computational complexity and memory usage compared to other decoding
algorithms for NB-LDPC codes.  We believe that our work for the
hardware-efficient algorithm will encourage researchers to explore
the use of NB-LDPC codes in emerging applications.

\bibliographystyle{IEEEtran}
\bibliography{xhchen}



\end{document}